\documentclass[conference]{IEEEtran}
\IEEEoverridecommandlockouts

\usepackage{cite}
\usepackage{amsmath,amssymb,amsfonts}
\usepackage{algorithmic}
\usepackage{graphicx}
\usepackage{textcomp}
\usepackage{multirow}
\usepackage{epstopdf}
\usepackage{subcaption}
\usepackage{tabularx}
\usepackage{color}
\usepackage{array}
\usepackage{booktabs}
\usepackage{stfloats}
\usepackage{url}
\usepackage{verbatim}
\usepackage{makecell}
\usepackage{multirow}
\usepackage{threeparttable}
\usepackage{subcaption} 
\usepackage{float}
\usepackage{xcolor}
\def\BibTeX{{\rm B\kern-.05em{\sc i\kern-.025em b}\kern-.08em
    T\kern-.1667em\lower.7ex\hbox{E}\kern-.125emX}}
\usepackage[letterpaper,top=0.73in,bottom=1.09in,left=0.63in,right=0.63in]{geometry}

\usepackage[absolute]{textpos}

\begin{document}
	
\begin{textblock}{13}(1.39,0.35)
	\noindent Y. Guo, T. Zhang, S. Sun, M. Tao, and R. Gao, ``Measurement and analysis of scattering from building surfaces at millimeter-wave frequency," \textit{2025 IEEE Wireless Communications and Networking Conference Workshops (WCNC Wkshps)}, Milan, Italy, 2025.
\end{textblock}

\title{Measurement and Analysis of Scattering From Building Surfaces at Millimeter-Wave Frequency\\
}

\setlength{\columnsep}{0.241in}

\author{
    \IEEEauthorblockN{Yulu Guo$^{1}$, Tongjia Zhang$^{1}$, Shu Sun$^1$, Meixia Tao$^1$, Ruifeng Gao$^{2, 3}$}
    \IEEEauthorblockA{$^1$ Department of Electronic Engineering, Shanghai Jiao Tong University, Shanghai 200240, China}
    \IEEEauthorblockA{$^2$ School of Transportation and Civil Engineering, Nantong University, Nantong 226019, China}
    \IEEEauthorblockA{$^3$ Nantong Research Institute for Advanced Communication Technologies, Nantong 226019, China}
 }

\maketitle

\begin{abstract}
In future air-to-ground integrated networks, the scattering effects from ground-based scatterers, such as buildings, cannot be neglected in millimeter-wave and higher frequency bands, and have a significant impact on channel characteristics. However, current scattering measurement studies primarily focus on single incident angles within the incident plane, leading to insufficient characterization of scattering properties. In this paper, we present scattering measurements conducted at 28 GHz on various real-world building surfaces with multiple incident angles and three-dimensional (3D) receiving angles. The measured data are analyzed in conjunction with parameterized scattering models in ray tracing and numerical simulations. Results indicate that for millimeter-wave channel modeling near building surfaces, it is crucial to account not only for surface materials but also for the scattering properties of the building surfaces with respect to the incident angle and receiving positions in 3D space.
\end{abstract}


\section{Introduction}
The effect of scatterers in electromagnetic wave propagation is a fundamental issue in electromagnetic field theory and also a crucial aspect of channel modeling in low-altitude wireless networks where complex scatterer clusters formed by ground buildings play a key role in the underlying physical channel\cite{Khawaja2019CST}. In the vision of future air-to-ground integrated wireless communication networks, the high frequency and wide bandwidth of millimeter waves can meet the data rate requirements for high-throughput mobile applications \cite{Khawaja2017VTC}, for instance, enabling unmanned aerial vehicles to hover at favorable positions near buildings to maintain high-speed connections with users \cite{Rupasinghe16GLOBECOM}.
The relatively large path loss and reduced wavelength of millimeter waves render link quality highly sensitive to building surface characteristics, particularly their geometric configurations and material compositions. Consequently, scattering mechanisms should be incorporated into channel modeling frameworks for millimeter-wave and higher-frequencies \cite{Katwe24}.

In scattering propagation modeling, two main categories of models are currently employed. One category includes the beckmann-kirchhoff (BK) model \cite{Beckmann1987} and the effective roughness (ER) model \cite{Esposti07TAP}, both of which are based on optical ray theory and surface roughness considerations. The second category consists of the radar cross section model \cite{Rappaport2002, Ju2019ICC}, which is particularly effective in predicting received power in far-field scattering scenarios between the transmitter (Tx) and receiver (Rx) \cite{Rappaport2002}. Derived from the ER model, the directive model and backscattering lobe model \cite{Esposti07TAP} provide useful descriptions of scattering patterns for surfaces with varying roughness. 
Based on these models, the authors in\cite{Guo2024WCNC} measured the scattering patterns of three indoor materials at 180 GHz with an incident angle of 30°. Ray tracing simulations revealed that scattering significantly impacts the small-scale characteristics of the indoor channel. In \cite{Garcia2016Access}, various indoor materials were parameterized between 40 GHz and 60 GHz, and it was verified that once the model parameters for a specific material are determined, the scattering model for that material can be directly applied to different scenarios without further tuning. The work \cite{Ren20IJAP} and \cite{Tian2019Access} proposed a more comprehensive process for tuning scattering model parameters, with measurements conducted between 40 GHz and 50 GHz on marble and granite surfaces at an incident angle of 30°. The results confirmed that the tuning process is applicable to materials with different surface irregularities and inherent properties at millimeter-wave frequencies.

Although these studies have measured multiple materials across various frequency bands, for specific real-world building surfaces in outdoor scenarios, the measurement process is often constrained by practical limitations, requiring the determination of initial scattering model parameters with a limited number of measurement points. Furthermore, most of the aforementioned studies designed measurements based on a single incident angle and received power only in the incident plane, without adequately considering the variation in scattering properties at different incident angles and three-dimensional (3D) positions. In this paper, we outline the basic scattering mechanisms and propose a method for calculating initial scattering model parameters based on surface roughness at small incident angles. We then present measurement campaigns conducted on different building surfaces at 28 GHz. Moreover, we analyze the variations in received power and scattering patterns across multiple incident angles and 3D Rx positions by performing ray tracing and numerical simulations for different scattering models and material surfaces. 
The scattering properties across various surfaces can be synthesized and extended to complex scenarios, including multiple interacting surfaces or surfaces of distinct material types.

\section{Basic Scattering Mechanisms}
\subsection{Diffuse Scattering Process}
According to the conservation of energy, it is generally assumed that the specular reflection component on a rough surface undergoes scattering loss compared to that on a smooth surface of the same material \cite{Rappaport2002,Ju2019ICC}. In other words, the radiation energy of this component is redistributed into the scattering process and can be modeled by introducing a loss factor $R$ into the reflection coefficient $\Gamma$\cite{Rappaport2002} as follows
\begin{equation}
    \Gamma_{\text{rough}} = R\cdot\Gamma,
    \label{eq2}
\end{equation}
where $\Gamma_{\text{rough}}$ is the new reflection coefficient the for rough surface and $R$ is given by\cite{Boithias1987Radio}
\begin{equation}
    R = \exp\left[ -8\left(\frac{\pi h_{\text{rms}}\cos{\theta_i}}{\lambda}\right)^2 \right]I_0\left[8\left(\frac{\pi h_{\text{rms}}\cos{\theta_i}}{\lambda}\right)^2\right],
    \label{eq3}
\end{equation}
where $h_{\text{rms}}$ is the standard deviation of the surface protuberance height about the mean height, $\theta_i$ is the incident angle, $\lambda$ is the wavelength, and $I_0$ is the Bessel function of the first kind and zero order.
It should be noted that equations (\ref{eq2}) and (\ref{eq3}) are based on the assumption of small incident angles of the BK model. When the incident wave strikes the rough surface at a small grazing angle, $R$ may not adequately represent the scattering losses in such cases\cite{Ju2019ICC}. 

If the field intensity of the electromagnetic wave incident on a smooth surface  is $\bar{E}_i$, the field intensity norm $|\bar{E}_r|$ of the specular reflected wave and the field intensity norm $|\bar{E}_t|$ of the transmitted electromagnetic wave can be calculated respectively as $|\bar{E}_r|=\Gamma\cdot|\bar{E}_i|$ and $|\bar{E}_t|=T\cdot|\bar{E}_i|$,
where the reflection coefficient $\Gamma$ and transmission coefficient $T$ can be determined by the Fresnel principle. When considering the roughness of the incident surface, we can define that the scattering coefficient $S$ satisfies $|\bar{E}_s|=S\cdot|\bar{E}_i|$ \cite{Esposti07TAP},
where $|\bar{E}_s|$ is the field intensity norm of the scattered wave and we can derive the following equation by conservation of energy\cite{Esposti07TAP}
\begin{equation}
    \Gamma_{\text{rough}}^2+S^2+T^2=1.
    \label{eq7}
\end{equation}

\subsection{Method for Calculating Initial Scatting Coefficient $S$}
Under comprehensive measurement conditions, the scattering coefficient of a specific material can be directly determined through measurement. This is typically done by designing experiments to measure the material's $\Gamma_\text{rough}$ and $T$, and then combining these results with (\ref{eq7}) to obtain an initial value for the scattering coefficient $S$ \cite{Ren20IJAP}. However, when considering the scattering characteristics of a specific building wall, there are practical challenges in directly measuring the scattering coefficient. The unique materials and spatial structures used in the construction of buildings make it difficult to accurately obtain the measured values of $\Gamma_\text{rough}$ and $T$. To address this issue, since the transmission coefficient is nearly unaffected by the characteristics of wall irregularities \cite{Esposti07TAP, Esposti1999VTC}, we can use the wall's roughness and reflection coefficient to determine the theoretical value of the scattering coefficient $S$ as follows
\begin{equation}
    S=\sqrt{\left(1-R^2\right)\cdot\Gamma^2}.
    \label{eq8}
\end{equation}
Using the scattering coefficient $S$, the scattered field intensity in various directions can be predicted, and $S^2$ can denote the proportion of the incident power on the wall element which is scattered in all directions\cite{Esposti07TAP}.


\subsection{Diffuse Scattering Models}
\begin{figure}[!t]
    \centering   
    \includegraphics[width = 0.7\columnwidth]
    {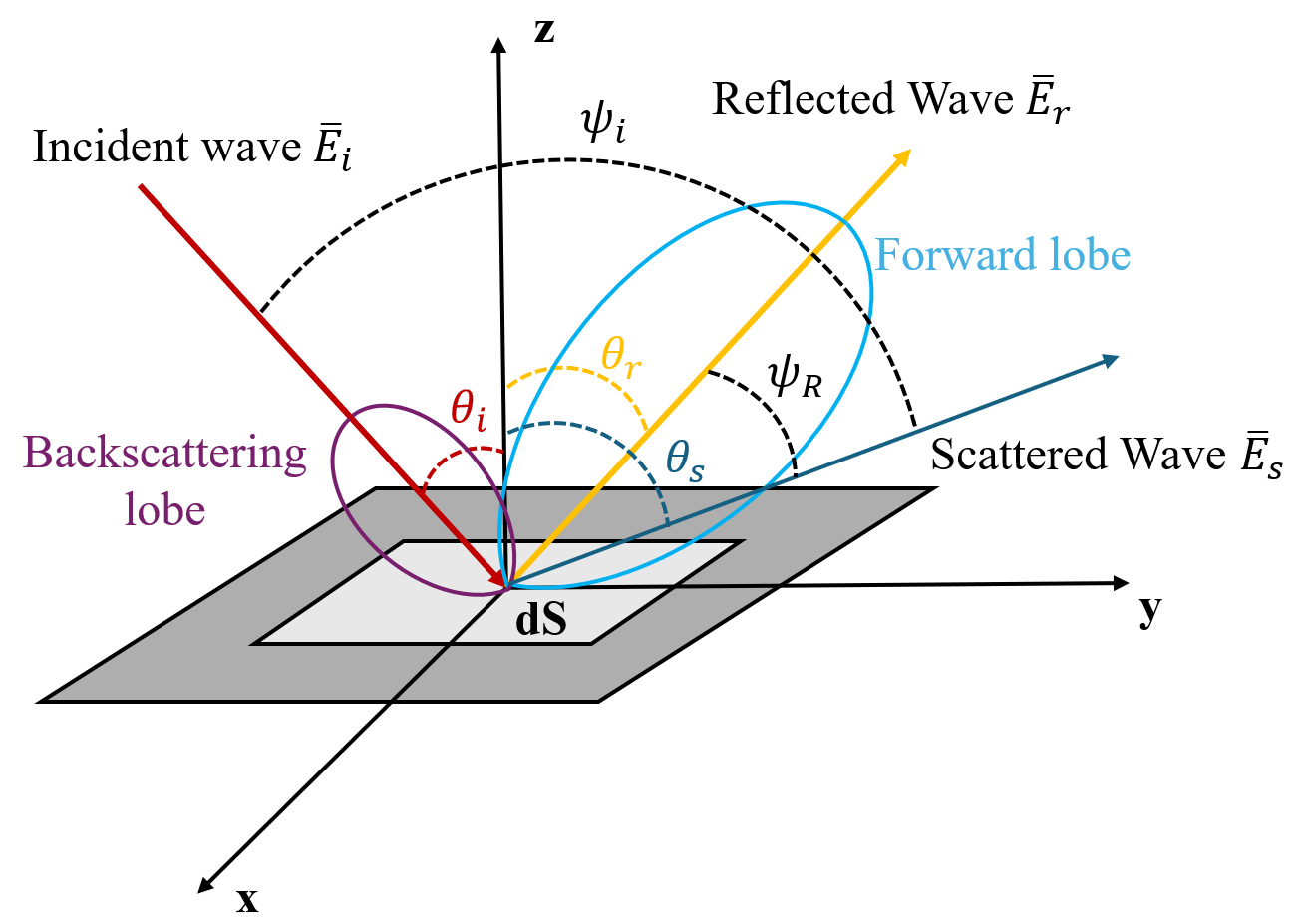}
    \caption{Illustration of scattering lobes.}   
    \label{fig1}
\end{figure}
In Fig. \ref{fig1}, we illustrate the scattering lobe patterns of an incident wave impinging on a rough surface element $\text{dS}$. The directive model and the backscattering lobe model are currently widely used for analyzing such scenarios.
\subsubsection{Directive model (single-lobe model)}
Full-wave simulations in \cite{Yi23JSTSP, Xie2022Radio} have shown that the scattering power is more concentrated in the specular reflection direction. Therefore, it is necessary to incorporate directional considerations into the scattering model, as illustrated by the directive model below
\begin{equation}
        \left| \bar{E}_s \right|^2 = \left( \frac{S K}{r_i \cdot r_s} \right)^2 \cdot\frac{l \cos \theta_i}{F_{\alpha_R}} \cdot \left( \frac{1 + \cos \psi_R}{2} \right)^{\alpha_R},
\end{equation}
where $F_{\alpha_R}$$=$$\int_{-\pi/2}^{\pi/2} \left( \frac{1 + \cos \psi_R}{2} \right)^{\alpha_R}$$ \sin \theta_s d\theta_s$, $\alpha_R$ $=$$ 1, 2, \ldots, 10$, $\theta_s$ is the angle between the scattering and the surface normal directions, $K$ is a constant depending on the amplitude of the incident wave and is given by $K=\sqrt{60P_tG_t}$ where $P_t$ and $G_t$ are transmitted power and antenna gain. $r_i$ and $r_s$ are distances between the Tx antenna and the incident point, and between the incident point and Rx antenna, respectively. $\psi_R$ is the angle between the scattering and the specular reflection directions, $l$ denotes the length of the surface, and $\alpha_R$ is the width factor of the scattering lobe. The larger the value of $\alpha_R$, the narrower the scattering lobe, resulting in better directional concentration of the scattered wave.

\subsubsection{Backscattering lobe model (dual-lobe model)}
For rougher and more irregular surfaces, the directivity of the scattering pattern becomes more complex, with an increased scattering component near the incident direction. To model this effect, the scattering concentrated in the backward direction was incorporated into the single-lobe model, thus allowing it to accurately characterize surfaces with more pronounced protrusions. The modified model is expressed as
\begin{equation}
    \begin{aligned}
        \left| \bar{E}_s \right|^2=&\left( \frac{S K}{r_i \cdot r_s} \right)^2 \cdot\frac{l \cos \theta_i}{F_{\alpha_R, \alpha_i}} \cdot  \Bigg[ \Lambda\cdot \left( \frac{1 + \cos \psi_R}{2} \right)^{\alpha_R} \\
        &+ (1 - \Lambda)\cdot \left( \frac{1 + \cos \psi_i}{2} \right)^{\alpha_i} \Bigg],\\
    \end{aligned}
\end{equation}
where $F_{\alpha_R, \alpha_i} $$=$$ \Lambda\cdot\int_{-\pi/2}^{\pi/2}\left( \frac{1 + \cos \psi_R}{2} \right)^{\alpha_R}\sin \theta_s d\theta_s+ (1 - \Lambda)\cdot \int_{-\pi/2}^{\pi/2}\left(\frac{1 + \cos \psi_i}{2} \right)^{\alpha_i} \sin \theta_s d\theta_s$, $\alpha_R$, $ \alpha_i $$=$$ 1, 2, \ldots, 10$, $\Lambda \in [0, 1]$, $\psi_i$ is the angle between the scattering and the incident directions, $\alpha_i$ is the width factor of the scattering lobe like $\alpha_R$, $\Lambda$ is a proportional factor that controls the amplitude of the forward scattering and the backscattering lobes. 

\section{Measurement Campaign}
\subsection{Measurement System}
\begin{figure}[!t]
    \centering   
    \includegraphics[width=0.9\columnwidth]
    {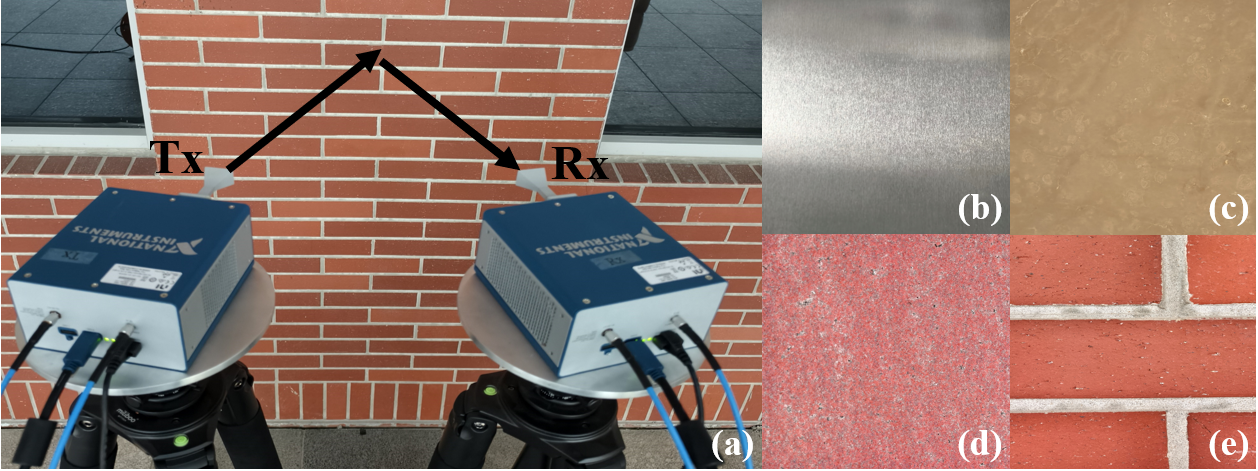}
    \caption{(a) Measurements scenario. (b) Metal sheet. (c) Marble wall. (d) Smooth wall. (e) Rough wall.}            
    \label{fig2}
\end{figure}
\begin{figure}[!t]
    \centering   
    \includegraphics[width=0.7\columnwidth]
    {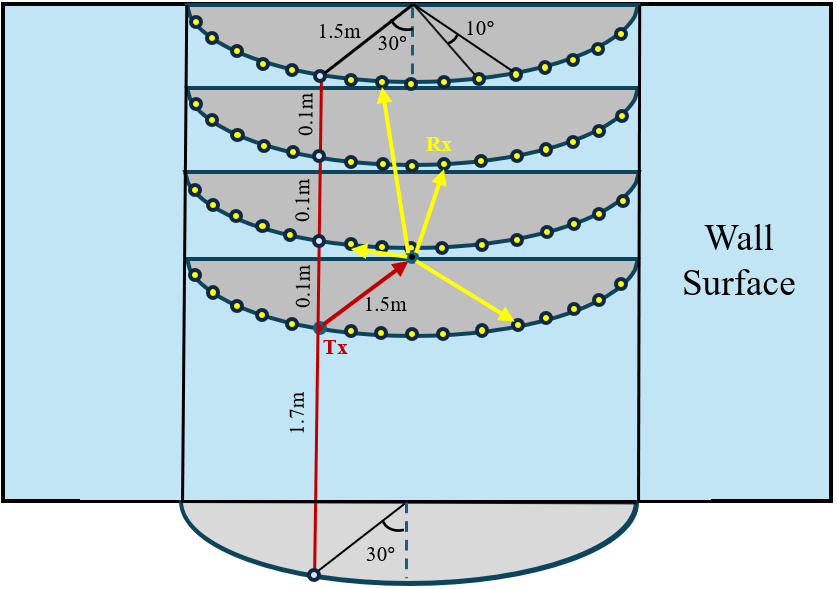}
    \caption{Measurements diagram for the rough wall. 
}            
    \label{fig3}
\end{figure}

We utilized a millimeter-wave time domain channel sounder system based on National Instruments hardware to conduct the measurement campaign. This channel sounder leverages commercial radio frequency (RF) front-ends for up and down conversion, employs field programmable gate arrays, and operates in a superheterodyne architecture with flexible intermediate frequency and local oscillator frequencies. The system's baseband signal is a Zadoff-Chu sequence with a length of 65,535, and the channel impulse response and power delay profile are obtained after correlation processing at the receiving end, with a maximum delay resolution of 650 ps.
The synchronization between the Tx and Rx is achieved by two rubidium clocks that have been pre-synchronized. 

\subsection{Measurement Scenario}

The scattering patterns of several typical building surfaces were measured at 28 GHz on the Minhang campus of Shanghai Jiao Tong University. 
Existing channel characterizations around the 28 GHz band (e.g., \cite{Ko17TWC,Rappaport17TAP,Jiang20ITVT}) have provided foundational propagation models, while scattering phenomena at this band remain insufficiently quantified for many real-world materials.
For the specific scenario measurements, two vertically polarized horn antennas, each with a gain of 15 dBi and a half-power beamwidth of 23°, were mounted on tripods at a height of 1.7 m above the ground at the RF ends, as shown in Fig. \ref{fig2}(a). The transmit power before entering the Tx antenna was 10 dBm, and the bandwidth was 2 GHz. 
The measured surfaces included a metal sheet and typical building wall surfaces on the campus, as shown in Figs. \ref{fig2}(b)-\ref{fig2}(e), with physical parameters listed in Table \ref{table1}.

During each measurement, the Tx antenna was fixed in position and orientation, located 1.5 meters from the center of the wall surface, while the Rx antenna was moved around a semicircle with a 1.5-meter radius with a step of 10 degrees. The center of the wall surface was in the far field of both Tx and Rx antennas \cite{Sun25Comm}.
In the measurement campaign on the rough wall surface with $\theta_i=30$°, we kept Tx fixed at a height of 1.7 m and set the heights of Rx to 1.7 m, 1.8 m, 1.9 m, and 2.0 m, moving the Rx along a semicylindrical surface,  as shown in Fig. \ref{fig3}. The Rx antenna was always directed towards the center of the wall surface. The received power at each position was recorded for subsequent analysis.
It is worth noting that during the measurements, the arrival times of the specular reflection and scattering components were close, and the dominant propagation paths were similar, making it difficult to accurately distinguish between the specular reflection power and the scattering power based solely on arrival delays. Therefore, all multipath power received within the maximum excess delay of 5 ns is considered as the total received power, where the paths with a path length difference within 1.5 m relative to the path of maximum power are regarded as contributing to the primary scattering and reflection.

\begin{table}[!]
\centering
\caption{\textsc{Parameters of Different Surfaces}}
\label{parameter_table}
\begin{tabular}{ccccc}
\toprule
Material  &  & $\epsilon_{r}$& $h_{\text{rms}}$ (mm) & Thickness (cm)\\ \midrule
Metal sheet &     & $6.0$    & 0.170  & 0.3 \\
Marble wall &     & $6.2 $   & 0.216  & 15\\
Smooth wall &    & $5.8$    & 0.445  & 25\\
Rough wall  &      & $10.5$   & 0.715   & 32\\ \bottomrule
\end{tabular}
\label{table1}
\end{table}

\begin{table*}[ht]
\centering
\caption{\textsc{Parameterization of Diffuse Scattering Model}}
\label{parameter_table}
\begin{tabular}{ccccccccc}
\toprule
Material  & Incident Angle $\theta_i$(°) & Diffuse Scattering Model & Theoretical $S$ & Best-Fit Range of $S$ & $\alpha_R$ & $\alpha_i$&$\Lambda$ & Minimum FVU \\ \midrule
metal sheet & 30     & single-lobe   & 0.1099 & $0.05 \sim 0.15$ & 4 &  - &  -& 0.2925 \\
marble wall & 30     & single-lobe   & 0.1409 & $0.10 \sim 0.20$ & 4 &  - &  -& 0.3544 \\
smooth wall & 30     & single-lobe   & 0.2815 & $0.20 \sim 0.30$ & 4 &  - &  -& 0.2223 \\
rough wall  & 20     & dual-lobe     & 0.7462 & $0.65 \sim 0.75$ &  1& 10 & 0.1& 0.6273 \\
rough wall  & 30     & dual-lobe     & 0.6174 & $0.55 \sim 0.65$ & 1 & 10 &  0.2& 0.5673 \\
rough wall  & 40     & dual-lobe     & 0.4791 & $0.45 \sim 0.55$ &  1&10   & 0.3& 0.4434\\\bottomrule
\end{tabular}
\label{table2}
\end{table*}

\section{Measurement and Simulation Results}
\subsection{Measured Data and Ray Tracing Results}

\subsubsection{Ray tracing settings}
\begin{figure}[!t]
    \centering   
    \includegraphics[width=0.7\columnwidth]{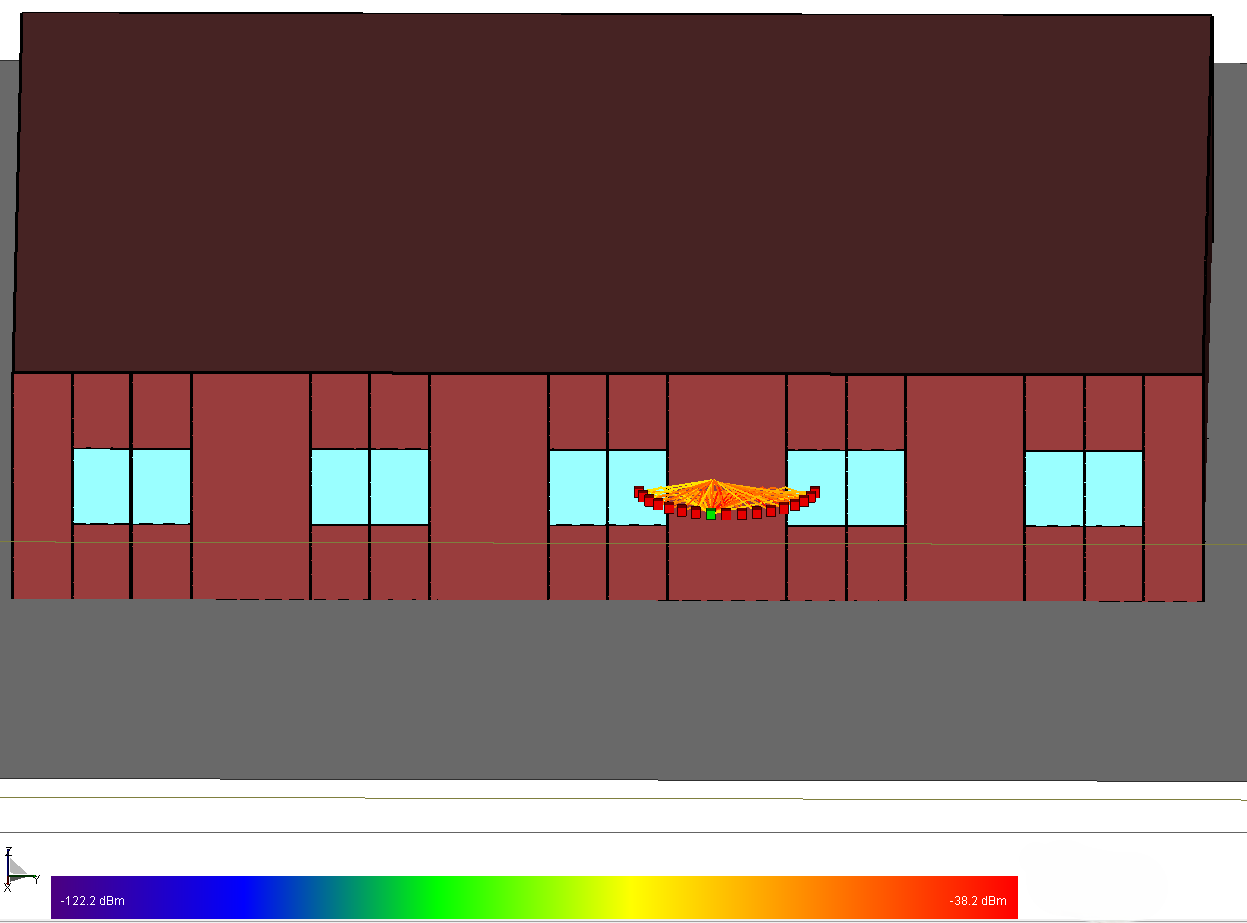}
    \caption{Ray tracing in the Wireless Insite.}            
    \label{fig4}
\end{figure}

In the ray tracing software Wireless Insite, we model the scenes of different building surfaces. Fig. \ref{fig4} shows the simulation of a rough wall surface, where the green point indicates the position of the Tx antenna and the red points indicate the different positions of the Rx antenna. The antenna positions and configuration parameters are the same as those used in the actual measurement. 
In Fig. \ref{fig4}, only the one-hop reflection path and one-hop scattering path are selected which contribute to the majority of the total received power. Other paths outside the wall, as well as multi-hop reflection and scattering paths, contribute very little to the total received power. These paths have power levels more than 35 dB below the main path power and are considered as noise, primarily originating from the ground or other surrounding surfaces.

\subsubsection{Parameterization of diffuse scattering model}
We employ both single-lobe and dual-lobe scattering models to analyze the scattering characteristics of the measured building surface materials. 
The scattering coefficient $S$ has the most significant effect on the scattering pattern, while $\alpha_R$ and $\alpha_i$ have a relatively weak influence \cite{Garcia2016Access}. The reference value of $\Lambda$ should decrease as the incident angle increases \cite{Ju2019ICC}. Based on the data recorded in Table \ref{table1}, the theoretical scattering coefficient for the surface under the unique incident angle can be determined by combining (\ref{eq2}), (\ref{eq3}), and (\ref{eq8}). which is applied in the ray tracing software as the initial setting for the scattering coefficient. Other parameters are adjusted to minimize the fraction of variance unexplained (FVU) whose expression is as follows
\begin{equation}
\text{FVU} = \sqrt{\frac{\sum\limits_{i=1}^{M} \left| P_{\text{measured, $i$}}(dB) - P_{\text{simulated, $i$}}(dB) \right|^2}{\sum\limits_{i=1}^{M} \left| P_{\text{measured, $i$}}(dB) - \bar{P}_{\text{measured}}(dB) \right|^2}},
\end{equation}
where $M$ is the total number of the reception positions, $P_{\text{measured, $i$}}$ and $P_{\text{simulated, $i$}}$ are respectively measured power and simulated power at the $i$-th position, and $\bar{P}_{\text{measured}}$ is the mean received power of all the positions. After other parameters are preliminarily determined, the scattering coefficient $S$ is adjusted around the initial value and the above process is repeated until the optimal scattering model parameters are obtained, corresponding to the minimum FVU. The parameterizations for different surfaces are provided in Table \ref{table2}.

\subsubsection{Comparison of results}

\begin{figure*}[!t]
    \centering
    \begin{subfigure}{0.25\textwidth}
        \centering
        \includegraphics[width=\linewidth]{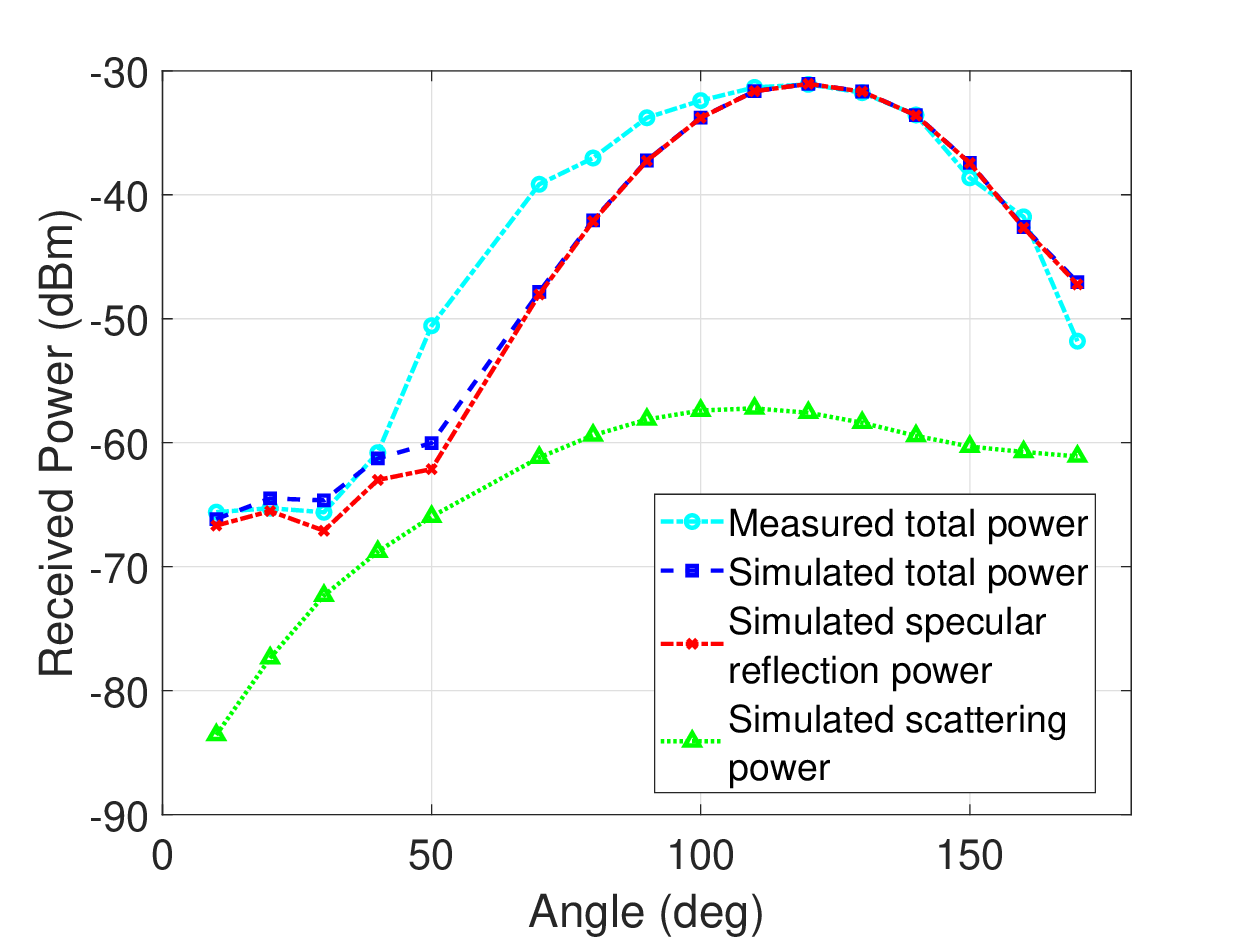}
        \caption{metal sheet, $\theta_i=30$°.}
   
    \end{subfigure}%
    \hfill
    \begin{subfigure}{0.25\textwidth}
        \centering
        \includegraphics[width=\linewidth]{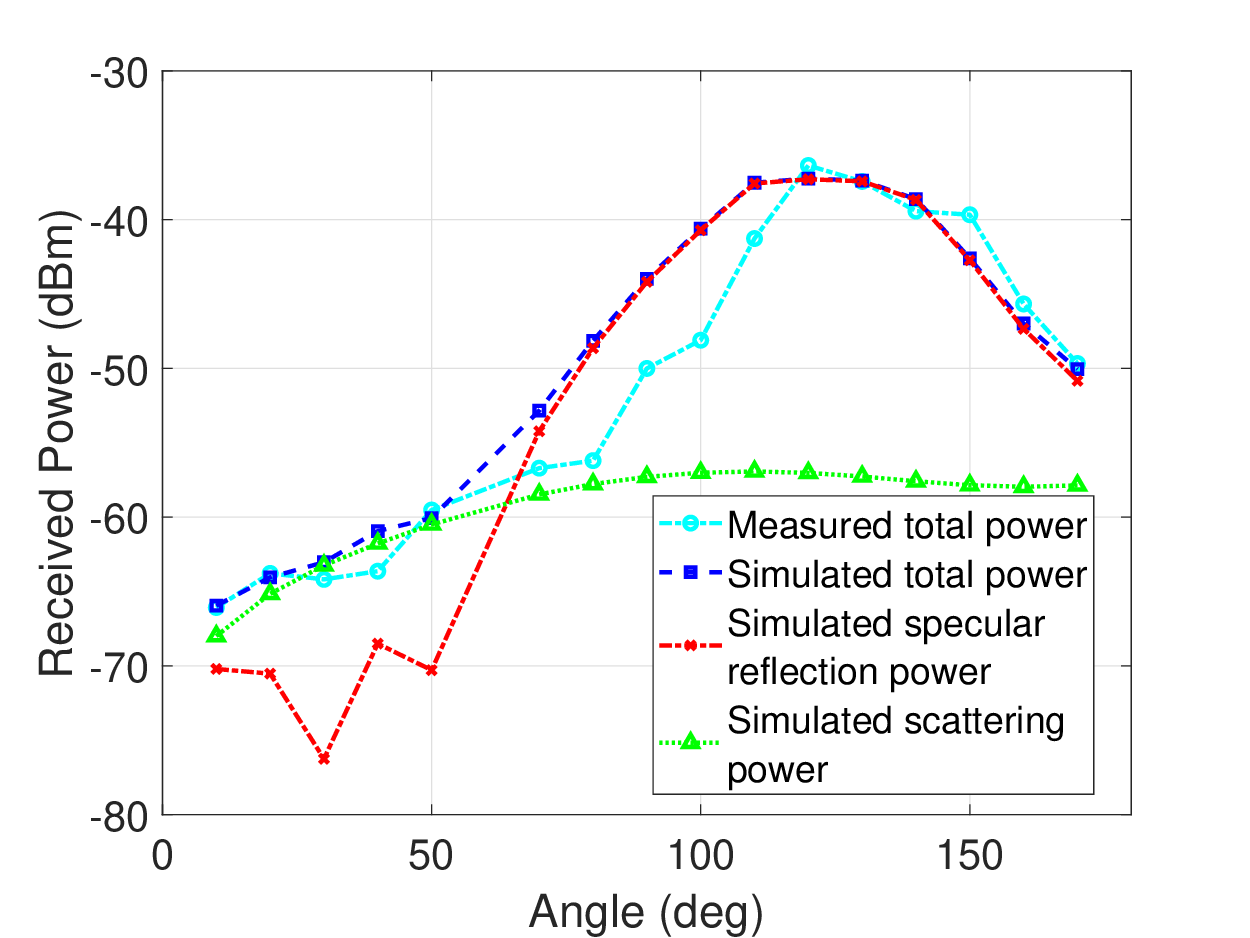}
        \caption{marble wall, $\theta_i=30$°.}

    \end{subfigure}%
    \hfill
    \begin{subfigure}{0.25\textwidth}
        \centering
        \includegraphics[width=\linewidth]{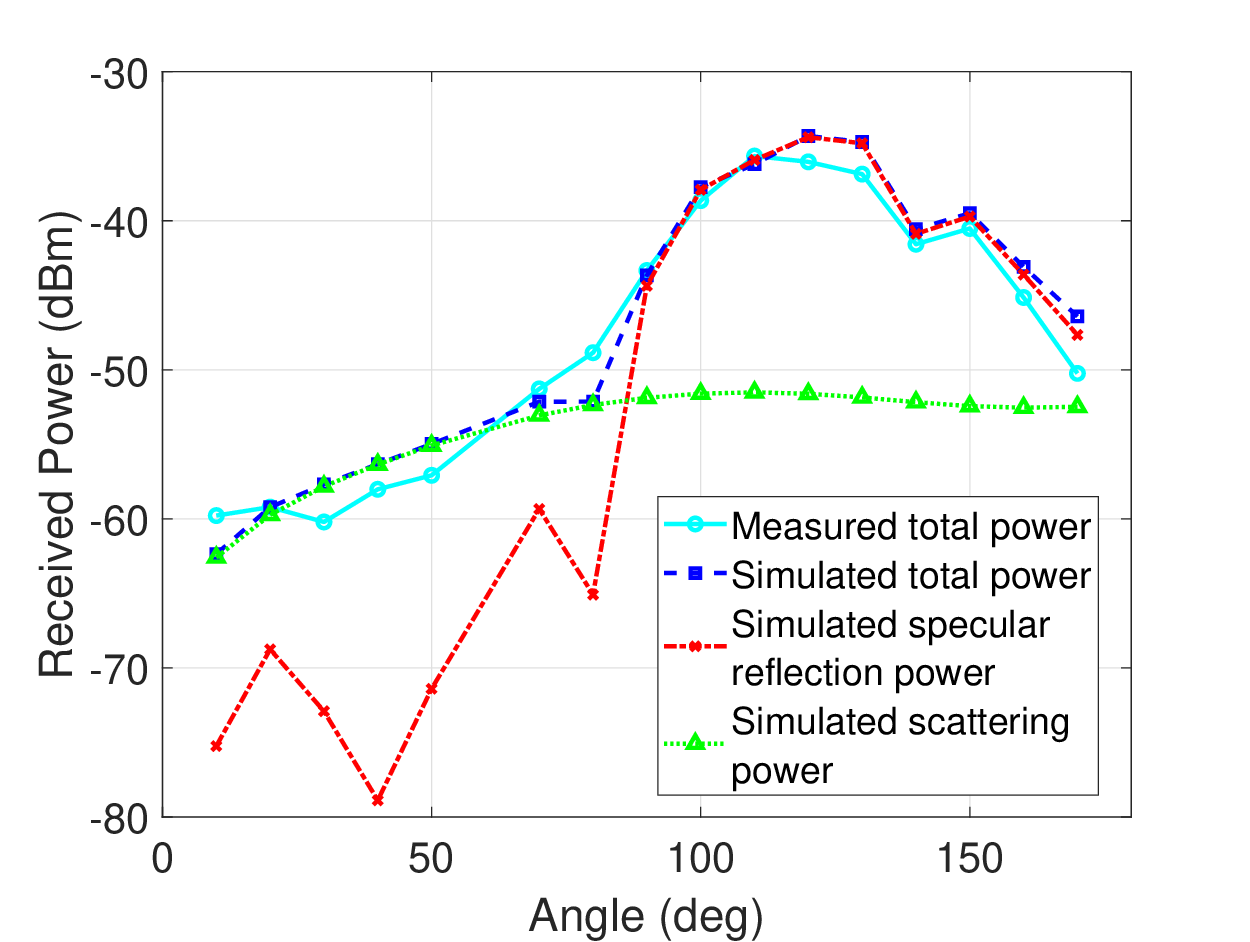}
        \caption{smooth wall, $\theta_i=30$°.}

    \end{subfigure}%
    \hfill
    \begin{subfigure}{0.25\textwidth}
        \centering
        \includegraphics[width=\linewidth]{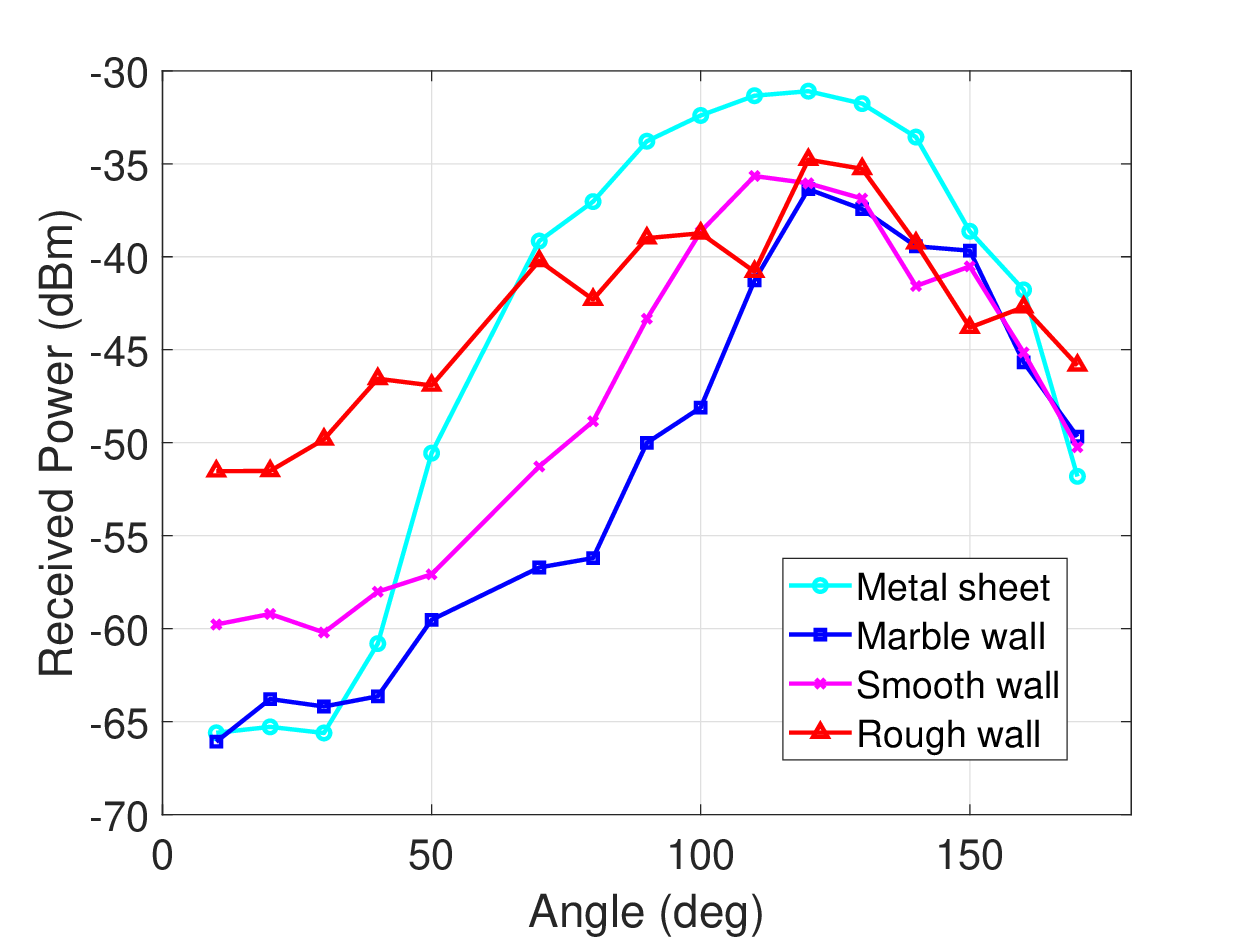}
        \caption{four material surfaces.}
   
    \end{subfigure}
    \begin{subfigure}{0.25\textwidth}
        \centering
        \includegraphics[width=\linewidth]{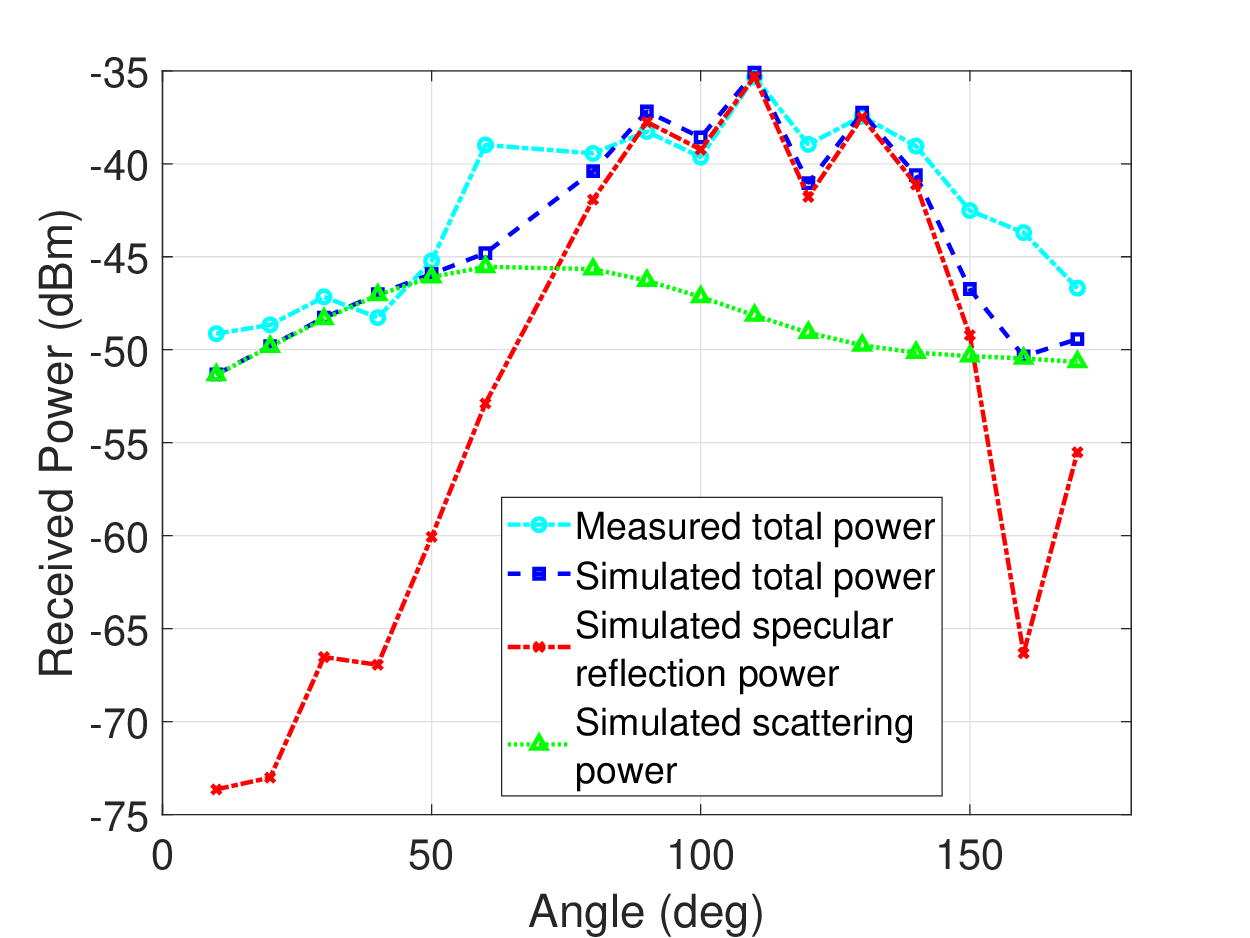}
        \caption{rough wall, $\theta_i=20$°.}

    \end{subfigure}%
    \hfill
    \begin{subfigure}{0.25\textwidth}
        \centering
        \includegraphics[width=\linewidth]{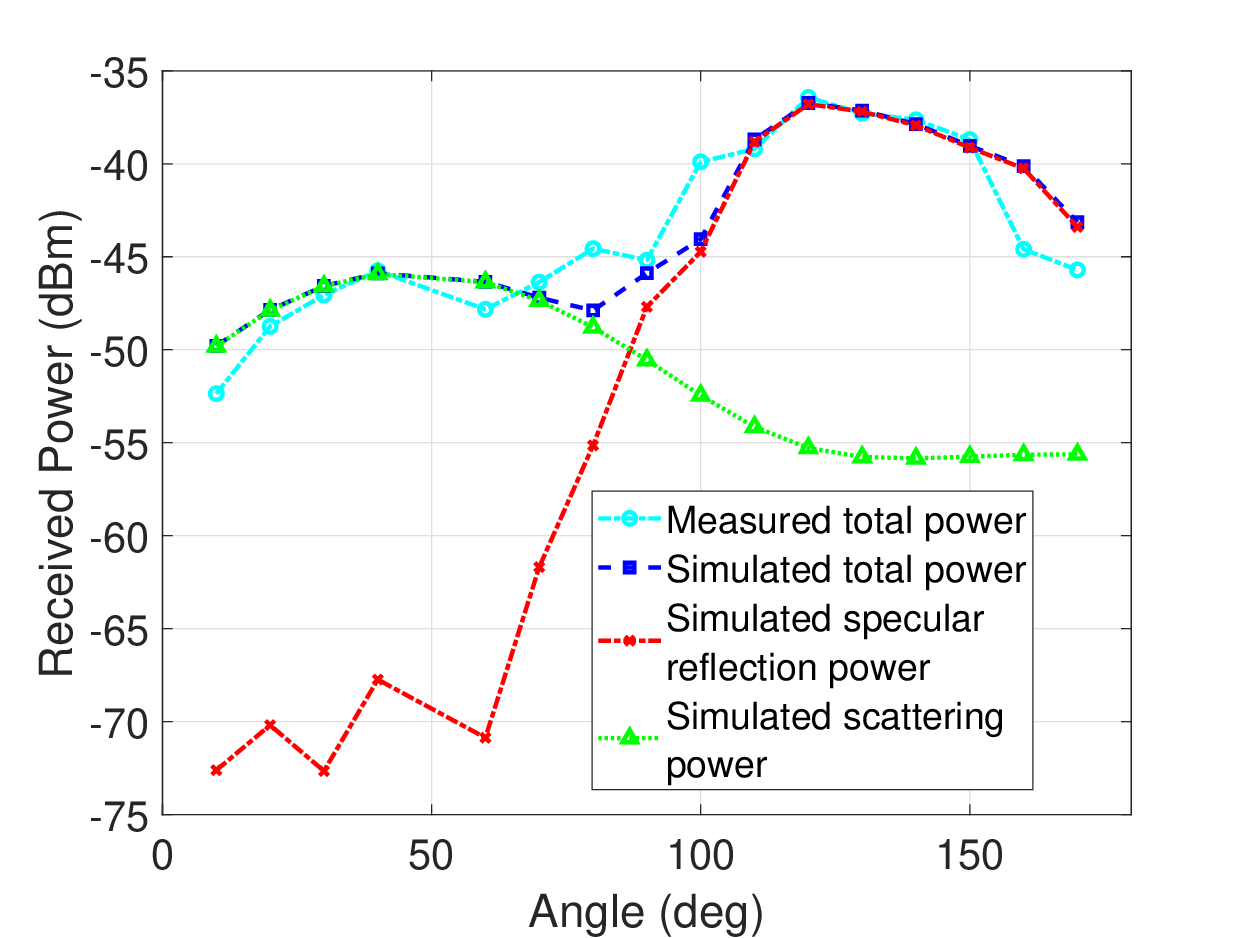}
        \caption{rough wall, $\theta_i=30$°.}

    \end{subfigure}%
    \hfill
    \begin{subfigure}{0.25\textwidth}
        \centering
        \includegraphics[width=\linewidth]{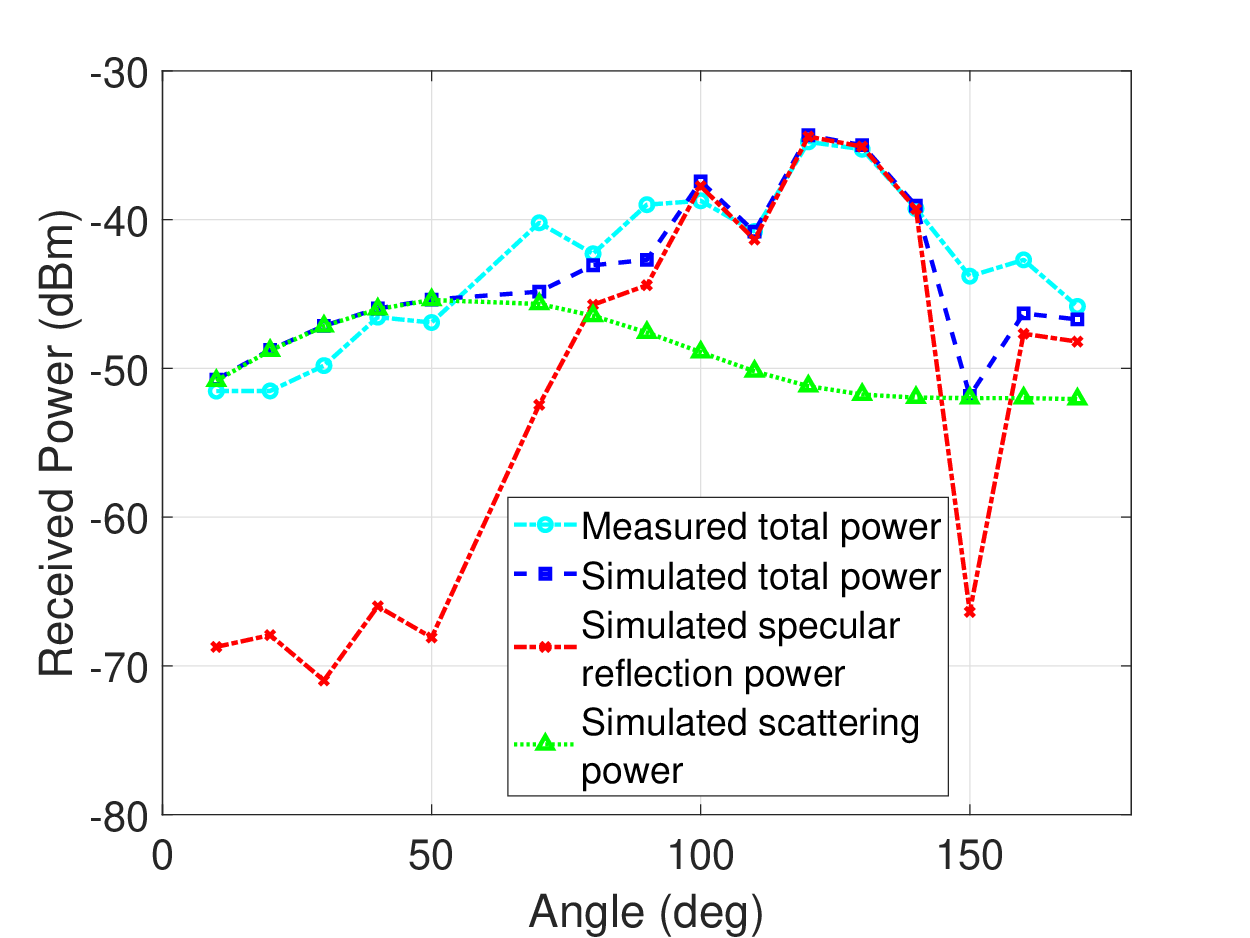}
        \caption{rough wall, $\theta_i=40$°.}
  
    \end{subfigure}%
    \hfill
    \begin{subfigure}{0.25\textwidth}
        \centering
        \includegraphics[width=\linewidth]{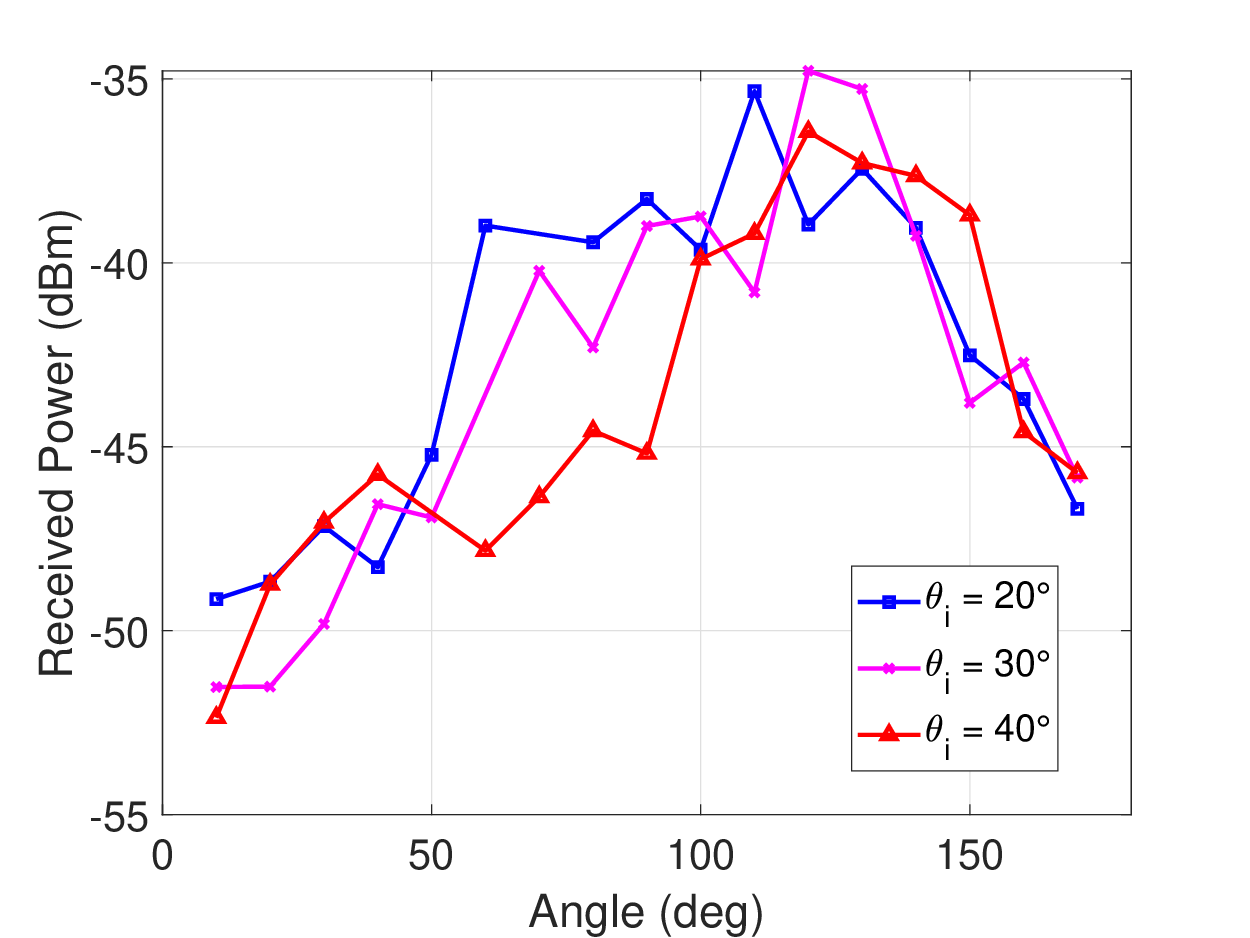}
        \caption{three incident angles.}
 
    \end{subfigure}
    \caption{(a)-(c), (f) Measurements and ray tracing for the four material surfaces at $\theta_i=30$°. (d) Comparison of measured power for different surfaces at $\theta_i=30$°. (e), (g) Measurements and ray tracing for the rough wall surface at $\theta_i=20$° and $\theta_i=40$°, respectively. (h) Comparison of measured power at different $\theta_i$ for the rough wall surface.}
    \label{fig5}
\end{figure*}

In Figs. \ref{fig5}(a)-\ref{fig5}(c), and \ref{fig5}(f), we compare the measured total received power and the simulated total received power on four material surfaces at $\theta_i = 30$°. The specular reflection and scattering paths are selected based on the ray tracing results to record the specular reflection power and scattering power in the simulation, and the measured total received power for four materials at $\theta_i = 30$° is directly compared in Fig. \ref{fig5}(d). For the metal sheet in Fig. \ref{fig5}(a), the specular reflection component dominates the received power at all received angles, reaching its maximum in the specular reflection direction, while the scattering power remains low. This strong reflection characteristic makes metal materials commonly used as reference materials for determining the reflection component. Compared with the marble wall in Fig. \ref{fig5}(b) and the smooth wall in Fig. \ref{fig5}(c), the specular reflection component’s power is concentrated near the specular reflection direction, sharply decreasing as the angle deviates to the incident side, while the scattering power proportion increasing and becoming the dominant component of the total received power. Also shown in Fig. \ref{fig5}(d), due to the increased roughness of the wall material, which leads to a larger $S$, more power is radiated in various scattering directions and shows an increase in the received power from scattering paths in the incident side, especially for the rough wall. In this case, the more suitable dual-lobe model was used to fit the measured data.

Figs. \ref{fig5}(e)-\ref{fig5}(h) demonstrate the variations in backscattering patterns induced by changes in the incident angle within the small-angle regime on rough surfaces. As the incident angle decreases, the scattering loss governed by (\ref{eq2}) and (\ref{eq3}) increases. This is manifested by a corresponding rise in the scattering coefficient \( S \) and a slight increase in the scattered power at small angles, a phenomenon further corroborated by the subsequent results in Fig. \ref{fig7}.
For the dual-lobe model, the incident direction and specular reflection direction gradually separate as $\theta_i$ increases. From Figs. \ref{fig5}(e)-\ref{fig5}(g), a noticeable increase in the backscattering component is observed with the decreasing best-fit scattering coefficient and increasing $\Lambda$ in Table \ref{table2}, meaning the reduction of scattering effects. It is also reflected in Fig. \ref{fig5}(h) as a reduction in the received power between the forward and backward directions.
\subsubsection{Parameterization with measured data corresponding to 3D Rx space}
\begin{figure}[!t]
    \centering
    \includegraphics[width=0.8\columnwidth]{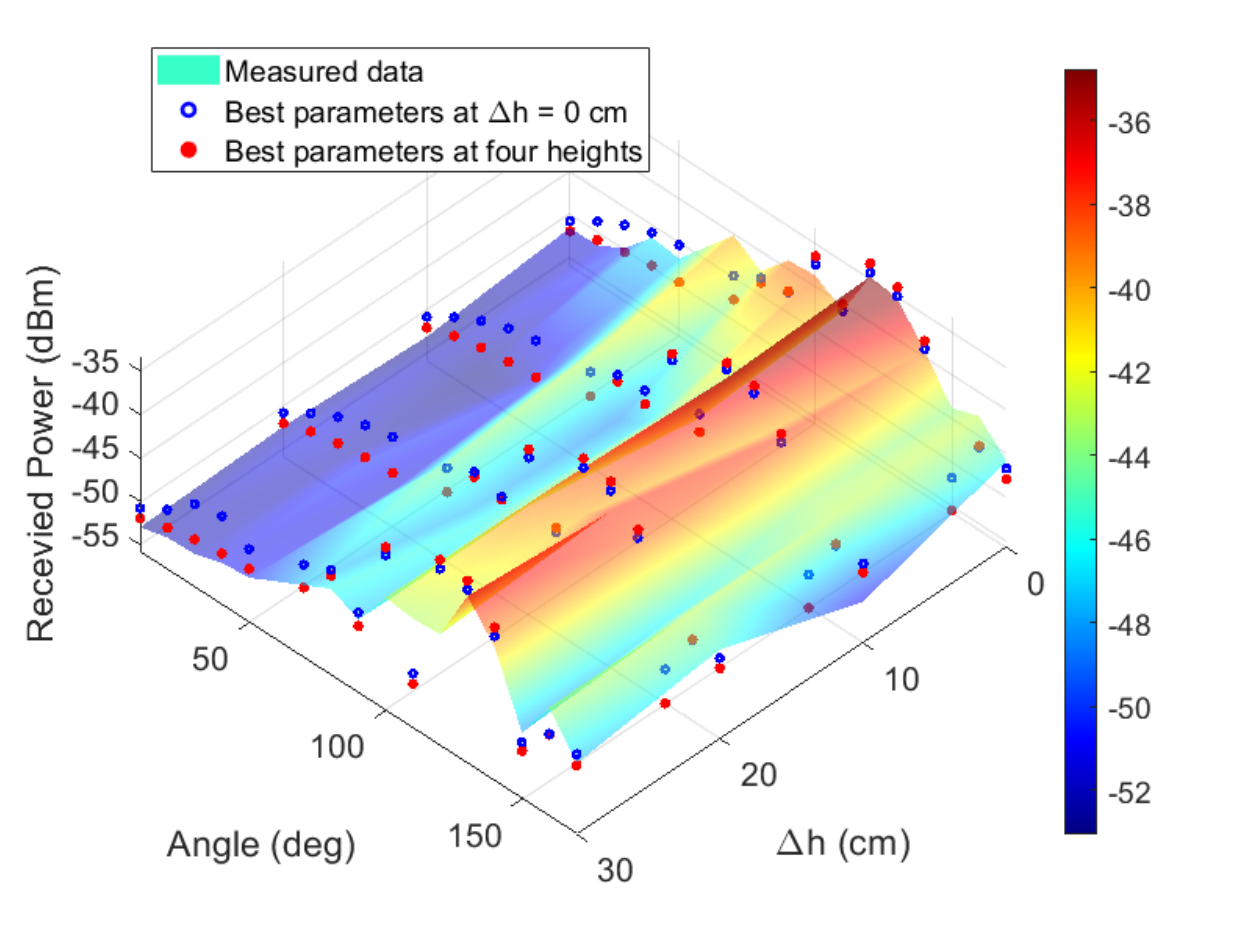}
    \caption{Recevied power of different positions.}      
    \label{fig6}
\end{figure}
The scattering is actually distributed throughout the space in front of the incident surface, rather than being confined to the incident plane. In traditional parameterization processes \cite{Guo2024WCNC,Garcia2016Access,Ren20IJAP,Tian2019Access}, only the ray tracing results within the incident plane are considered for fitting the measured data to determine the scattering model parameters, as demonstrated previously. This fitting process overlooks the changes caused by the spatial distribution of scattering power. To address this, under rough wall measurement conditions, we varied the height of Rx by $\Delta h$, integrated the obtained data, and then applied the fitting process described above. 
The received power at various 3D positions is illustrated in Fig. \ref{fig6}, where the blue circles represent the results of the ray tracing with the best parameters using data only in the incident plane ($\Delta h=0$ cm), the red dots represent the results of the ray tracing with the best parameters at four Rx heights ($\Delta h=0, 10, 20, 30$ cm), and the colored surface represents the measured data.
As the Rx height increases, it can be observed that the scattering model determined based on the incident plane ($S$ = 0.60, $\alpha_R=1$, $\alpha_i=10$, $\Lambda=0.2$) incorrectly predicts the actual backscattering power outside the plane ( FVU = 0.6742). On the other hand, by fitting the data obtained from multiple stereoscopic positions, the scattering model ($S$ = 0.42, $\alpha_R=6$, $\alpha_i=4$, $\Lambda=0.2$) can partially correct this error while still maintaining a good fitting performance within the incident plane ( FVU = 0.6259). This indicates that for a more accurate and detailed analysis of scattering effects, it is necessary to consider the scattering effects in the stereoscopic space, within an acceptable level of complexity, to jointly determine the actual scattering model parameters.
\subsection{Numerical Simulation}
\begin{figure}[!t]
    \centering
    \begin{subfigure}[t]{\columnwidth}
        \centering
        \includegraphics[width=0.8\columnwidth]{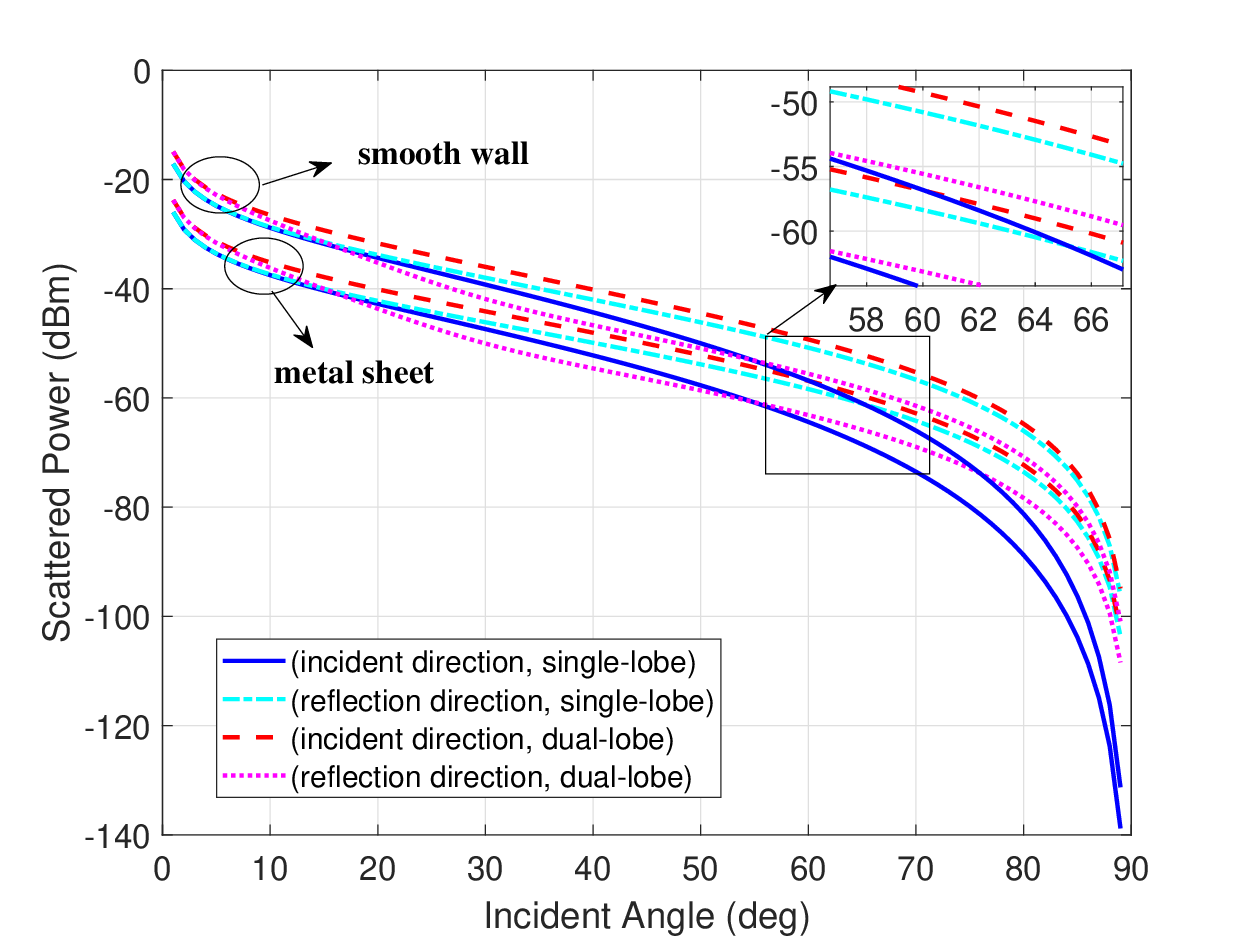}
        \caption{metal sheet and smooth wall.}        
    \end{subfigure}
    \begin{subfigure}[t]{\columnwidth}
        \centering
        \includegraphics[width=0.8\columnwidth]{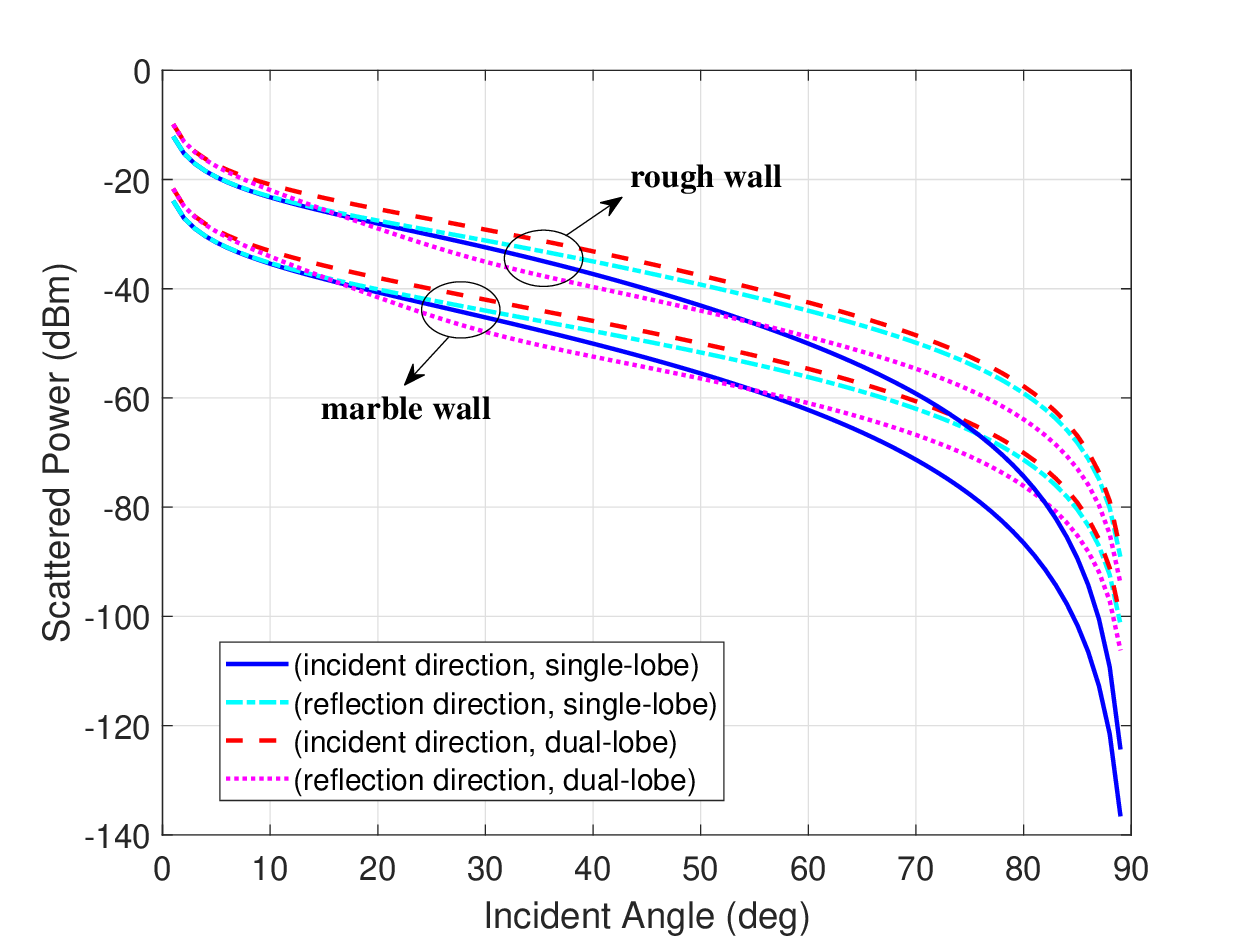}
        \caption{marble wall and rough wall.}       
    \end{subfigure}
    \caption{Numerical simulation results of scattered power. To simplify the calculation, the parameters other than $S$ are set as specified in Table \ref{table2}, and $\Lambda$ is uniformly set to 0.2.}
    \label{fig7}
\end{figure}

Due to the physical constraints and limitations of the measurement apparatus, it is not feasible to directly measure the power radiated from the surfaces at higher angular resolutions. Therefore, we employ numerical simulations to model the scattering characteristics at an incident area dS. The power in the incident and specular reflection directions is calculated using both single-lobe and dual-lobe models as \cite{Rappaport2002}
\begin{equation}
    \begin{aligned}
        P_r= \frac{G_r\lambda^2}{480\pi^2}\cdot\left| \bar{E}_s \right|^2,\\
    \end{aligned}
\end{equation}
where $P_r$ represent the received power at different directions. The calculation results for the four materials listed in Table \ref{table1} are presented in Fig. \ref{fig7}. When $\theta_i$ decreases, as shown in Fig. \ref{fig7}, the scattered power in both directions gradually increases. Notably, the received scattered power near the direction of the incident angle approaching the normal direction shows a significant increase, while for $\theta_i$ greater than 60°, the scattered power decreases rapidly. Within the $\theta_i$ range of 1° to 25°, the single-lobe model can be approximated by assuming equal scattered power in both directions. In contrast, for the dual-lobe model, the predicted power in both directions is primarily influenced by the value of $\Lambda$ and tends to converge as the angle decreases. Among the four materials shown in Figs. \ref{fig7}(a) and \ref{fig7}(b), the rough wall demonstrates the strongest scattering capability, while the metal plate exhibits the weakest scattering. The scattering power of the smooth wall and marble wall is nearly identical due to their similar roughness, which aligns with the measurement results presented in Figs. \ref{fig5}(b) and \ref{fig5}(c). 

\section{Conclusion}
In this study, we have analyzed the scattering mechanisms and determined the initial scattering coefficients for parameterization, which are influenced by surface roughness and reflection coefficients. We conducted real-world measurements at 28 GHz, and selected a metal sheet along with several typical building wall surfaces on the campus as test surfaces, obtaining received power data for various incident angles and 3D positions. We have applied the parameterized scattering models in ray tracing and numerical simulations, revealing clear trends in both scattering power and specular reflection power as surface roughness and positions in space varied. 
Specifically, the scattering parameters for real-world building surfaces should be determined collectively by the scattering power at different spatial positions under varying incident angles. 
These findings underscore the importance of considering scattering characteristics in channel modeling for the design of future millimeter-wave and higher-frequency communication networks, especially around complex building scatterers. 


\vspace{12pt}


\begin{thebibliography}{00}
\bibitem{Khawaja2019CST} W. Khawaja \emph{et al}., “A survey of air-to-ground propagation channel modeling for unmanned aerial vehicles," \emph{IEEE Commun. Surveys Tuts.}, vol. 21, no. 3, pp. 2361–2391, thirdquarter 2019.
\bibitem{Khawaja2017VTC} W. Khawaja \emph{et al}., “UAV air-to-ground channel characterization for mmWave systems,” in \emph{Proc. IEEE Veh. Technol. Conf.}, Toronto, ON, Canada, Sep. 2017, pp. 1–5.
\bibitem{Rupasinghe16GLOBECOM}N. Rupasinghe \emph{et al}., “Optimum hovering locations with angular domain user separation for cooperative UAV networks,” in \emph{Proc. IEEE Global Telecommun. Conf. }, 2016, pp. 1–6.
\bibitem{Katwe24} M. V. Katwe \emph{et al.}, “CmWave and sub-THz: Key radio enablers and complementary spectrum for 6G,” [Online]. Available: https://arxiv.org/pdf/2406.18391.
\bibitem{Beckmann1987}P. Beckmann \emph{et al}., \emph{The Scattering of Electromagnetic Waves From Rough Surfaces}. Norwood, MA: Artech House, 1987.
\bibitem{Esposti07TAP}V. Degli-Esposti \emph{et al}., “Measurement and modeling of scattering from buildings,” \emph{IEEE Trans. on Ant. and Prop.}, vol. 55, no. 1, pp. 143–153, Jan 2007.
\bibitem{Rappaport2002} T. S. Rappaport, \emph{Wireless Communications: Principles and Practice}, 2nd ed. Upper Saddle River, NJ: Prentice Hall, 2002.
\bibitem{Ju2019ICC}S. Ju \emph{et al.}, “Scattering mechanisms and modeling for terahertz wireless communications,” in \emph{Proc. IEEE Int. Conf. Commun. }, May 2019, pp. 1–7.
\bibitem{Guo2024WCNC}G. Guo \emph{et al}., "Diffuse scattering analysis of indoor propagation channel at terahertz frequency," \emph{Proc. IEEE Wireless Commun. Netw. Conf. }, Apr. 2024, pp. 1–6.
\bibitem{Garcia2016Access}J. Pascual-Garcia \emph{et al}., “On the importance of diffuse scattering model parameterization in indoor wireless channels at mmWave frequencies,” \emph{IEEE Access}, vol. 4, pp. 688-701, Feb. 2016.
\bibitem{Ren20IJAP}M.-H. Ren \emph{et al}., “Diffuse scattering directive model parameterization
 method for construction materials at mmWave frequencies,” \emph{Int. J. Antennas Propag.}, vol. 2020, pp. 1–9, Dec. 2020.
\bibitem{Tian2019Access}H. Tian \emph{et al}., “Effect level based parameterization method for diffuse scattering models at millimeter-wave frequencies,” \emph{IEEE Access}, vol. 7, pp. 93286–93293, 2019.
\bibitem{Boithias1987Radio} L. Boithias \emph{et al., Radio wave propagation.} McGraw-Hill NY, 1987.
\bibitem{Esposti1999VTC}V. Degli-Esposti and H. L. Bertoni, “Evaluation of the role of diffuse scattering in urban microcellular propagation,” in \emph{Proc. IEEE Veh. Technol. Conf. }, Amsterdam, The Netherlands, Sep. 1999, pp. 1392–1396.
\bibitem{Yi23JSTSP}H. Yi \emph{et al}., “Full-wave simulation and scattering modeling for terahertz communications,” \emph{IEEE J. Sel. Topics Signal Process.}, vol. 17, no. 4, pp. 1–16, Nov. 2023.
\bibitem{Xie2022Radio}P. Xie \emph{et al}., “Terahertz wave propagation characteristics on rough surfaces based on full-wave simulations,” \emph{Radio Sci.}, vol. 57, no. 6, 2022, Art. no. e2021RS007385.
\bibitem{Ko17TWC}J. Ko \emph{et al}., “Millimeter-wave channel measurements and analysis for statistical spatial channel model in in-building and urban environments at 28 GHz", \emph{IEEE Trans. Wireless Commun.}, vol. 16, no. 9, pp. 5853-5868, Sep. 2017.
\bibitem{Rappaport17TAP}T. S. Rappaport \emph{et al}., “Small-scale, local area, and transitional
millimeter wave propagation for 5G communications,” \emph{IEEE Trans. on Ant. and Prop.}, vol. 65, no. 12, pp. 6474-6490, Dec. 2017.
\bibitem{Jiang20ITVT}T. Jiang \emph{et al}., “The comparative study of S–V model between 3.5 and 28 GHz in indoor and outdoor scenarios", \emph{IEEE Trans. Veh. Technol.}, vol. 69, no. 3, pp. 2351-2364, Mar. 2020.

\bibitem{Sun25Comm} S. Sun \emph{et al}., “How to differentiate between near field and far field: Revisiting the Rayleigh distance," \emph{IEEE Commun. Magazine}, vol. 63, no. 1, pp. 22-28, Jan. 2025.


\end{thebibliography}
\end{document}